\documentclass[12pt,a4paper]{article}

\usepackage[margin=25mm]{geometry}
\usepackage{newtxtext,newtxmath}
\usepackage{microtype}
\usepackage{setspace}
\usepackage{multirow}
\usepackage{array}

\usepackage{graphicx}
\graphicspath{{/mnt/project/}}
\usepackage{caption}
\usepackage{subcaption}
\usepackage{float}
\usepackage{placeins}
\usepackage{pdflscape}
\usepackage{booktabs}
\usepackage[round,sort&compress]{natbib}

\newcommand{\keywords}[1]{%
  \par\addvspace{\baselineskip}\noindent\textbf{Keywords:} #1
}

\PassOptionsToPackage{hyphens}{url}
\usepackage[hidelinks]{hyperref}

\hypersetup{
  pdftitle={FLARE v2: A Recursive Framework for Program Comprehension Across Common Teaching Languages and Levels of Abstraction},
  pdfauthor={Justin Heath},
  pdfsubject={Computing education; semiotics; program comprehension}
}

\newcolumntype{P}[1]{>{\raggedright\arraybackslash}p{#1}}

\title{FLARE v2: A Recursive Framework for Program Comprehension Across Common Teaching Languages and Levels of Abstraction}
\author{%
Justin Heath\\
\small Independent Researcher\\
\small justinheath@outlook.com \;|\; https://jjgh.me
}
\date{\today}

\begin{document}

\maketitle


\begin{abstract}
\small
Building on the classroom framework in Heath et al. (2025), this paper proposes FLARE v2 as a recursive, semiotically informed account of how program meaning can be described across abstraction scales in common teaching languages. It reframes FLARE v1's tiers as one cycle: identify bounded elements (Receives, Sends, Effects, Shares), analyse bindings along two dimensions (Causal-Temporal and Communicative), and treat the bound set as a new element at the next scale. Causal-Temporal binding has three subtypes - Sequential, Branch, and Event - to distinguish user-authored control flow from event-driven control whose dispatch is hidden in the runtime. A Compositional Ladder visualises the same compositional move from blocks and statements through segments and systems.

FLARE v2 is scoped to imperative and event-driven environments typical of primary and lower-secondary curricula. Above the system layer, behaviour is increasingly shaped by interaction between code and operating context (scheduling, infrastructure, permissions, contracts, failures, platform policy). Here, the element-and-binding vocabulary remains a structural baseline, but continuity of explanation typically requires overlays that make environmental constraints explicit. Event binding and overlays serve a common pedagogical role - preserving coherent structural reasoning where key causal mechanisms are not fully visible in the authored artefact. OOP design reasoning, explicit concurrency models, distributed systems, and functional paradigms are treated as future extensions; implementation and evaluation are left for future work.
\end{abstract}

\clearpage

\section{Introduction}

Teaching programming requires teachers to help pupils make sense of program behaviour, yet many computing teachers are non-specialists. The Royal Society reported that many computing teachers felt under-supported and lacked specialist background in computer science \citep{royalsociety2017}, and a recent follow-up review found that only around a third of England’s computer science ITT recruitment target is being met and that just 40\% of computing teachers hold a relevant post-A-level qualification \citep{royalsociety2025}. In addition, much of a program's runtime behaviour is not directly visible in the source text. Sorva argues that, without explicit teaching about a notional machine, novices form incomplete and unstable understandings of how programs execute \citep{sorva2013notional}. Sentance and colleagues likewise emphasise the importance of explicit classroom talk about what code does, not only about writing it \citep{sentance2023programming}.

Program comprehension research has characterised code understanding as structural and compositional. Early work on procedural languages identified programming "plans" - stereotyped code patterns that experts recognise and use to solve families of problems \citep{soloway1984empirical}, as summarised in later reviews \citep{schulte2010block}. Pennington's studies showed that comprehending a program involves constructing both a control-flow oriented program model and a data-flow or situation model, and coordinating between them \citep{pennington1987comprehension,schulte2010block}. Schulte's Block Model synthesises this work into a matrix with hierarchical levels (atoms, blocks, relations, macro) and dimensions of understanding (program text, execution, function) \citep{schulte2008block}. These frameworks articulate key abstractions that underpin expert comprehension, but say less about how meanings at different scales - from individual blocks to whole systems - are related in classroom discourse. In primary settings, teachers therefore still need practical ways of talking with pupils about structure, flow and purpose. Sentance and colleagues highlight that students and teachers require explicit language for explaining how programs work and why they behave as they do \citep{sentance2023programming}.

FLARE~v1 \citep{heath2025flare} responded to this need by adapting the Block Model for primary classrooms as four tiers for talking about block-based programs: Blocks, Segments, Relationships, and Macro. The CEP~'25 study showed that these tiers helped teachers structure questioning and classroom dialogue, but also exposed limitations. Teachers reported that the Relationships and Macro tiers were difficult to apply in simple programs and, at times, confusing; classroom analysis suggested that the Relationships tier in particular combined several distinct mechanisms, including temporal ordering, causal triggering, and shared state \citep{heath2025flare}. Introducing all four tiers at once also placed demands on teachers' subject knowledge and working memory.

This paper presents FLARE~v2 as a recursive reformulation in which elements bind to form new elements at successive scales. Section~\ref{sec:scope} states the intended contribution and boundaries of the claim. Section~\ref{sec:diagnosis} summarises the constraints that emerged in the CEP~'25 work, before Sections~\ref{sec:semiotic} and~\ref{sec:generative-core} introduce the semiotic framing and the generative core of FLARE~v2. A later positioning section relates FLARE~v2 to established program comprehension models.

\section{Scope and Intended Contribution}
\label{sec:scope}

FLARE~v2 is proposed as a conceptual and analytic framework for reasoning about program meaning across several levels of abstraction. Its contribution is theoretical: a compact descriptive scheme for explaining how bounded program elements can be treated as units, related through bindings, and re-described at increasing scales. The framework aims to provide a shared classroom-facing vocabulary that can support teacher-led analysis, classroom dialogue, and reflective interpretation of code behaviour - particularly in contexts where execution mechanisms are partially hidden from learners.

The framework is intentionally descriptive rather than explanatory in a psychological sense. It does not claim to model the cognitive processes by which learners construct understanding, nor does it assert that novices naturally reason in its terms. Where empirical observations are reported - such as the Year~5 probe in Section~\ref{sec:semiotic} - these motivate design choices and suggest patterns worth investigating, but do not validate FLARE~v2 as a cognitive theory. Claims about pedagogical value are hypotheses for future study, not established effects.

The scope is bounded by programming paradigm and by where meaning can plausibly be treated as largely "in the artefact" for classroom analysis. FLARE~v2 has been developed for imperative and event-driven environments typical of primary and lower-secondary computing curricula, including block-based and hybrid languages. The element and binding vocabulary is intended to remain usable as scale increases from expressions and commands through segments and systems.

Beyond this, service- and ecosystem-level reasoning increasingly requires accounting for factors that sit outside the authored artefact, so continuity of explanation depends on explicitly pairing code-side structure with environmental constraints. In such settings, the behaviour of an otherwise stable code artefact may depend on non-local factors (runtime scheduling, infrastructure configuration, permissions, service contracts, versioning, latency, partial failure, and platform policy). This does not require abandoning a structural-compositional lens, but it does mean that a purely code-centred structural description is no longer sufficient on its own. In system-plus contexts, FLARE~v2 should be treated as describing the code-side composition, and paired with additional overlays that represent operational and socio-technical constraints.

Accordingly, this paper does not claim demonstrated applicability to object-oriented design reasoning, explicit concurrency models, distributed systems, or functional paradigms. Where such paradigms are discussed, they are treated as potential extensions rather than covered cases. In this sense, FLARE~v2 is best understood as a candidate analytic lens: its success should be judged by whether it proves generative for analysis, curriculum design, and subsequent empirical investigation.

\section{Legacy and Diagnosis: Constraints of FLARE v1}
\label{sec:diagnosis}

The CEP~'25 study demonstrated that FLARE~v1 supported structured discussion but also revealed structural and pedagogical limitations when used in primary classrooms \citep{heath2025flare}. These limitations stemmed from how the fixed tier hierarchy interacted with teacher subject knowledge, artefact scale, and cognitive load.

\subsection{Tier Granularity and Teacher Subject Knowledge}

Teachers engaged readily with the lower tiers - Blocks and Segments - because these mapped onto visible code structures. Difficulty arose with Relationships and Macro, which required reasoning about invisible mechanisms such as triggers and shared state.

Teacher~A stated: "I didn't understand half of them... There were elements that I was just totally baffled by."

Teacher~C similarly explained: "Because I've not had any coding experience... you're full, that's you at capacity."

These statements align with national evidence that many teachers lack specialist subject knowledge and feel under-supported in teaching the new curriculum \citep[p.~72]{royalsociety2017}. The Royal Society’s 2025 review similarly highlights “uneven provision and lack of suitably qualified subject specific teachers” in computing education and notes that computing teacher recruitment remains critically below government targets \citep{royalsociety2025}. They also sit alongside findings that non-specialist teachers' own subject knowledge and confidence shape how they interpret pupils' programming difficulties \citep{sentance2023programming}.

\subsection{Mismatch Between Tier Structure and Artefact Scale}

FLARE~v1 assumed that programs would exhibit meaningful structure at all four tiers. However, the block-based artefacts used in CEP~'25 - simple, event-driven scripts - rarely contained the long-range dependencies or global design features required for productive use of the Relationships and Macro tiers.

Teacher~B noted: "They weren't linking into segments and therefore it wasn't working... you've got two different segments that aren't linked together."

This pattern resonates with Schulte et al.'s distinction between "programming in the large" and "programming in the small", where abstractions designed for large, complex systems may not apply straightforwardly to short instructional programs \citep[p.~75]{schulte2010block}.

\subsection{Cognitive Load and Simultaneous Introduction}

Teachers reported that introducing all four tiers at once created cognitive load. Rather than functioning as interpretive lenses, the tiers sometimes became a checklist.

Teacher~A reflected: "Once I got going, it was like a snowball... but looking back at it, it was still overwhelming."

Teacher~C added: "I didn't have the capacity to develop the questions further... I didn't have time to sit back and reflect."

Classroom dialogue also showed that the Relationships tier conflated distinct mechanisms - temporal order, causal triggering, and state-sharing - leading to confusion when teachers attempted to apply it. In Black and Wiliam's terms, teachers need time to invest in practices that deepen learning rather than simply pushing through content: they warn that '"delivery" and "coverage" with poor understanding are pointless and can even be harmful' \citep[p.~144]{black1998assessment}.

Table~\ref{tab:v1v2mapping} summarises how the four FLARE~v1 tiers map onto v2's elements and bindings.


\begin{table}[ht]
\centering
\caption{Mapping FLARE v1 tiers to v2 elements and bindings}
\label{tab:v1v2mapping}
\small
\begin{tabular}{p{0.20\linewidth} p{0.40\linewidth} p{0.32\linewidth}}
\toprule
\textbf{FLARE v1 Tier} & \textbf{FLARE v2 Element} & \textbf{FLARE v2 Binding} \\
\midrule
Blocks         & Elements at block scale          & (none) \\
Segments       & Elements at segment scale        & Causal-Temporal within segments \\
Relationships  & (none)                           & Both binding dimensions \\
Macro          & Elements at system scale         & Purpose emerges from bindings \\
\bottomrule
\end{tabular}
\end{table}

These limitations prompted a reconceptualisation. Where FLARE~v1 was primarily a pedagogical scaffold - a set of tiers to organise classroom talk - FLARE~v2 aims to be a generative theoretical account of how program meaning can be described at increasing scale.

\section{Formalising the Pattern: Code as Semiotic System}
\label{sec:semiotic}

The limitations identified in FLARE~v1 indicate the need for a descriptive lens that separates mechanisms rather than bundling them, and supports progressive introduction. Teachers' natural reasoning about code focused on sequencing, triggering, conditions, and shared state, yet FLARE~v1 did not articulate these distinctions.

FLARE~v2 is designed to address these challenges by grounding the framework in a light semiotic footing. Classical structural linguistics emphasises that meaning in language arises from relations within a system and from the composition of bounded units into larger structures \citep{saussure1916course,hjelmslev1961prolegomena}. FLARE~v2 adopts only a minimal commitment from this tradition: that bounded units carry meaning, and that binding these units yields larger units that can be re-described. This perspective supports the recursive element-binding cycle developed in the subsequent sections.

\begin{figure}[H]
    \centering
    \includegraphics[width=0.9\linewidth]{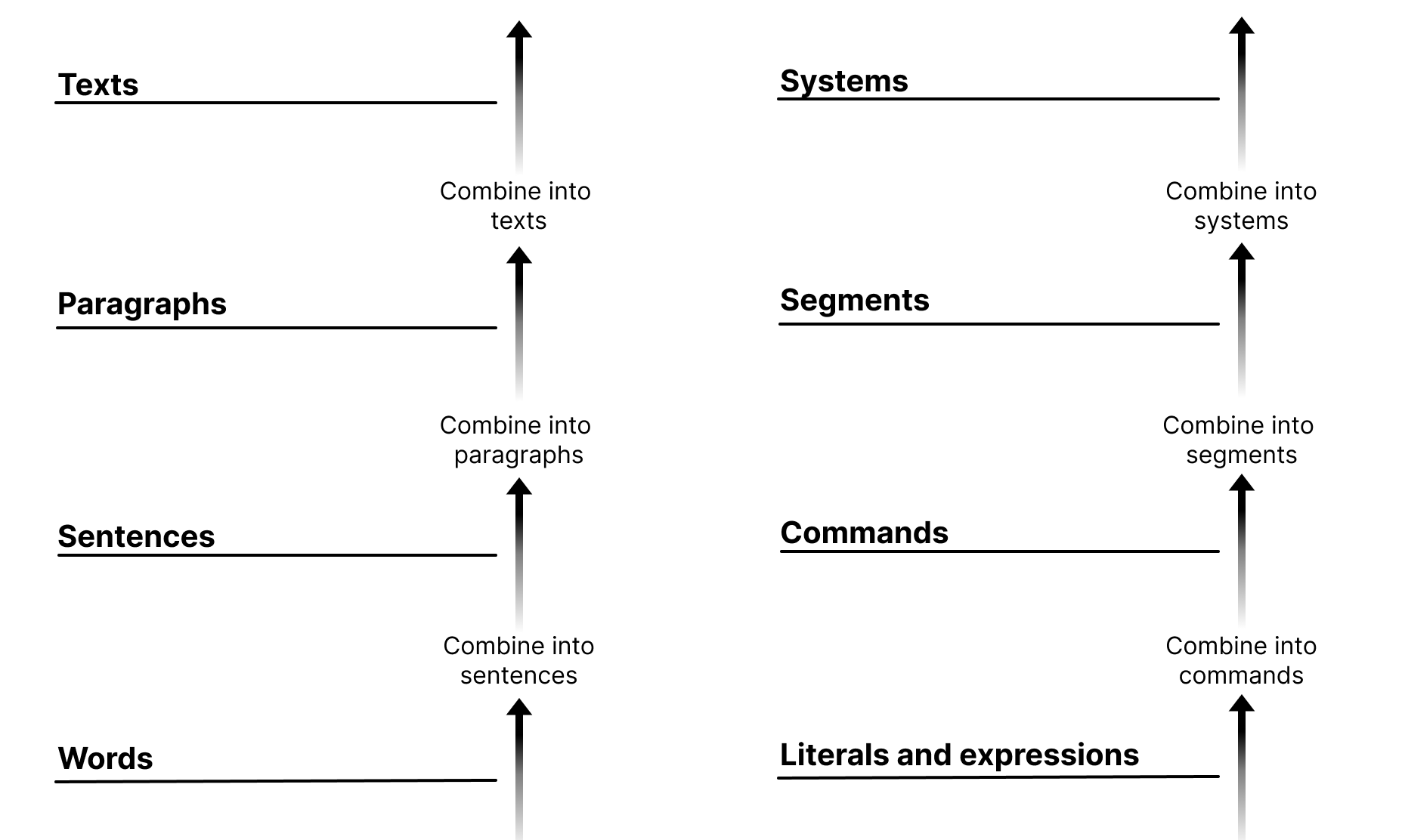}
    \caption{The Compositional Ladder: parallel compositional progressions in literacy and programming. Both domains build meaning through composition of bounded units into larger structures.}
    \label{fig:semiotic-ladder}
\end{figure}

Figure~\ref{fig:semiotic-ladder} illustrates how literacy and programming can be discussed using a common compositional metaphor. On the left, familiar literacy units (words, sentences, paragraphs, texts) combine through grammatical bindings. On the right, programming constructs (literals/expressions, commands, segments, systems) can be treated using an analogous compositional move. This parallel provides teachers with an accessible metaphor: just as sentences combine to form paragraphs that express complete ideas, commands combine to form segments that accomplish discrete computational tasks.

\subsection{Design origins}

The redesign of FLARE began from practice rather than theory. While using FLARE~v1 with non-specialist teachers in block-based environments that sit on a JavaScript event loop, certain programs repeatedly exposed gaps in the framework. Simple scripts that mix non-blocking timers with sequential commands often behave in ways that pupils and teachers find surprising. For example, a program that schedules \texttt{"morning"} to be printed after one second and then immediately prints \texttt{"Good"} will output \texttt{Good morning}, even though many learners predict the opposite. FLARE~v1 described structure at the level of Blocks, Segments, Relations and Macro, but it did not give teachers a precise way to talk about timing behaviour, non-blocking constructs, or the event loop that underpins them. The Relations tier in particular blurred temporal order, causal triggering and shared state.

In response, I set aside high-level classroom examples and worked downwards to smaller and smaller representations, asking what the most fundamental units of program structure might be. This led to a deliberately reductive exercise with a small RISC-style instruction set. Registers and memory locations were treated as "things", core instructions such as \texttt{MOV}, \texttt{ADD}, \texttt{SUB}, \texttt{JMP} and \texttt{CMP} as actions on those things, and branch instructions such as \texttt{BEQ} and \texttt{BNE} as choices after a comparison. The immediate goal was not to recover familiar control structures, but to see whether this reduction could separate the conflated concerns in FLARE~v1, particularly the mix of temporal sequencing and logical control that appeared in both the Segments and Relations tiers.

What emerged was a fundamental distinction in how control moves between instructions. Code that executes without any branch instruction - where the instruction pointer simply increments - differs categorically from code that requires a jump to a label. Sequential flow needs no branch; conditionals, loops, and function calls all require one. This binary provided a cleaner foundation than FLARE~v1's conflated Relations tier.

A further distinction became apparent when considering event-driven code. When a child writes an event handler such as \texttt{button.onClick(...)}, the handler is registered, but invocation is controlled by the runtime's event loop - an invisible \texttt{while(true)} that checks for events and dispatches handlers. Mechanically, in common event-loop runtimes this can be approximated as Branch inside a loop. Pedagogically, the dispatch mechanism is hidden from the learner, and so the handler is experienced as something that "just fires" when clicked. This motivated treating Event as a distinct subtype within Causal-Temporal binding in FLARE~v2.

At this point, a tension emerges between the visible code and the observed runtime behaviour. In sequential and branching code, the learner can usually explain why one element runs next by pointing to a structure they authored: the next line, an \texttt{if}, a loop, or a call. In event-driven code, that causal link is no longer fully legible in the authored artefact. The learner writes a handler and registers it, but the decision about \emph{when} it runs is made by a runtime dispatch mechanism that is typically invisible to them.

In FLARE~v1, this loss of visible causality contributed to confusion within the Relationships tier, where temporal order, triggering, and shared state were discussed without a clean separation of mechanisms. FLARE~v2 addresses this discontinuity by formalising the \textbf{Event} subtype within Causal-Temporal binding. The Event subtype functions as a \emph{pedagogical bridge}: it reifies hidden dispatch into a distinct, nameable structural relation. By treating the event connection as a binding, the framework allows learners to maintain a structural mode of reasoning (“this element is bound to that trigger”) even when the underlying mechanism of that binding lies in the runtime environment.

On this view, Event binding does not claim that event-driven behaviour is unconstrained or fundamentally different from branching control. Rather, it marks a boundary of visibility and provides continuity in the learner’s explanatory model until they are ready to examine how the runtime implements dispatch (for example, via an event loop, callback queue, or scheduler).

A subsequent exploratory classroom probe (n=26, ages 9-10) revealed an interesting pattern. When shown event-driven Scratch-style code and asked what would happen if a second event fired mid-execution, 77\% of pupils selected concurrent execution on the probe items, without prompts or worked examples in the session itself. These children had most recently encountered an event-handling metaphor two years earlier in Year~3, when we talked about indented code as "levels" and they proposed an event-handling metaphor, noting that multiple handlers could work simultaneously. They had received no further teaching on concurrency in the intervening years. This suggests that early metaphors may form lasting expectations about event-driven execution, and it motivates careful treatment of scheduling and concurrency as an overlay when moving beyond single-threaded classroom runtimes. Full methodology and results are reported in \citet{heath2025concurrency}.

Two design commitments for FLARE~v2 followed from this work. First, any construct that consistently behaves as a single unit in reasoning or classroom dialogue should be treated as an element, provided it can be described in terms of what it receives, sends, effects and shares. Second, the ways elements connect should be separated into two orthogonal dimensions: Causal-Temporal binding (how control moves, with Sequential, Branch, and Event subtypes) and Communicative binding (how data flows). Only after this design work did I encounter Pennington's distinction between control-flow and data-flow representations \citep{pennington1987comprehension}. The alignment between that work and the binding scheme described here is therefore a convergence that helps situate FLARE~v2 within existing theory, rather than the origin of the framework itself.

\FloatBarrier
\section{The Generative Core of FLARE v2}
\label{sec:generative-core}

\begin{figure}[htbp]
    \centering
    \includegraphics[width=0.75\linewidth]{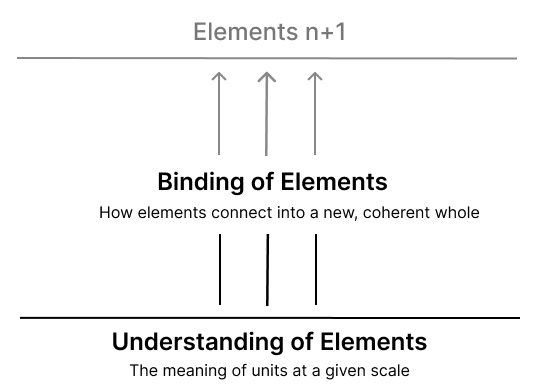}
    \caption{The generative core of FLARE~v2: elements at scale $n$ bind to form new elements at scale $n+1$. The same descriptive questions recur across levels; at higher layers additional overlays are often required.}
    \label{fig:generative-core}
\end{figure}

The Compositional Ladder and framing developed in Sections~\ref{sec:diagnosis} and~\ref{sec:semiotic} suggest a simple generative move that recurs across scales: elements bind to form new elements, which can themselves be treated as units in further bindings. Figure~\ref{fig:generative-core} visualises this recursive cycle, and FLARE~v2 formalises this move in two parts:

\begin{enumerate}
  \item a description of how individual elements can be expressed through four observable properties; and
  \item a set of binding dimensions that describe how elements combine to yield new bounded units.
\end{enumerate}

Together these form a recursive element-binding-element cycle that can be applied from blocks and statements through segments and systems, and extended to higher layers with additional overlays where required.

\subsection{Elements and their four properties}

FLARE~v2 uses the term \textit{element} for any bounded unit of code that can be treated as a single meaningful entity in classroom discussion. As summarised in Table~\ref{tab:v1v2mapping}, constructs that appeared as separate tiers in FLARE~v1 (for example, Blocks and Segments) are treated in v2 as elements at different scales but described using the same four properties.

Depending on context, an element may be a block, statement, function, script, subsystem, or larger component. Across the scales considered in this paper, the same four observable properties are used:

\begin{itemize}
\item \textbf{Receives}: inputs or triggers that cross the element boundary, such as events, parameters, sensor readings, inbound values, or user inputs.
\item \textbf{Sends}: outputs that cross the boundary, such as return values, emitted events, messages, or hardware commands.
\item \textbf{Effects}: internal state changes treated as confined within the element boundary, such as updates to local variables or private data structures.
\item \textbf{Shares}: any cross-boundary state or resource that the element reads or writes, such as global variables, configuration values, device state, shared hardware pins, shared objects, or external services.
\end{itemize}

The distinction between \textit{Effects} and \textit{Shares} is crucial. \textbf{Effects} are treated as private for inter-element reasoning. Any change that can be observed outside the element - because it modifies shared state, escapes via shared references, or interacts with hardware or external resources - is analysed as a \textbf{Share}. Transient messages or one-off calls that pass values between elements are treated as Sends, not Shares. Shares are reserved for persistent cross-boundary state or resources that can be read or written by multiple elements over time.

An important consequence of this distinction is that \textbf{Effects are not ordinarily treated as participating in bindings between elements}. Because Effects are analysed as private, they do not create inter-element connections. Inter-element Communicative bindings arise through Receives, Sends, and Shares. Effects matter for understanding what happens inside an element, but they do not, by definition, constitute a cross-boundary connection.

Table~\ref{tab:effects-shares} illustrates this distinction across different scales.


\begin{table}[h!]
\centering
\caption{Effects and Shares across scales}
\label{tab:effects-shares}
\small
\begin{tabular}{
P{0.18\linewidth}
P{0.36\linewidth}
P{0.36\linewidth}
}
\toprule
\textbf{Scale} & \textbf{Effects (private state)} & \textbf{Shares (cross-boundary state)} \\
\midrule
Single block
  & Updates a local counter used only inside the handler.
  & Writes to a global score variable or hardware pin. \\
Segment (function / handler)
  & Manipulates a private list or calculation scratch space accessible only within the segment.
  & Reads or writes a global threshold value or updates a device such as a microcontroller LED grid. \\
System
  & Maintains an internal scheduling index or mode flag used only within the system.
  & Interacts with shared configuration, logs to a cloud service, or changes hardware output. \\
\bottomrule
\end{tabular}
\end{table}

Because the same four properties can be used at each scale in the covered scope, teachers can move between discussing a single block and a whole program using stable questions: "What does this element receive?", "What does it send?", "What does it change inside itself?", and "What does it share?" The vocabulary remains consistent even as the grain size changes.

\begin{figure}[htbp]
    \centering
    \includegraphics[width=0.75\linewidth]{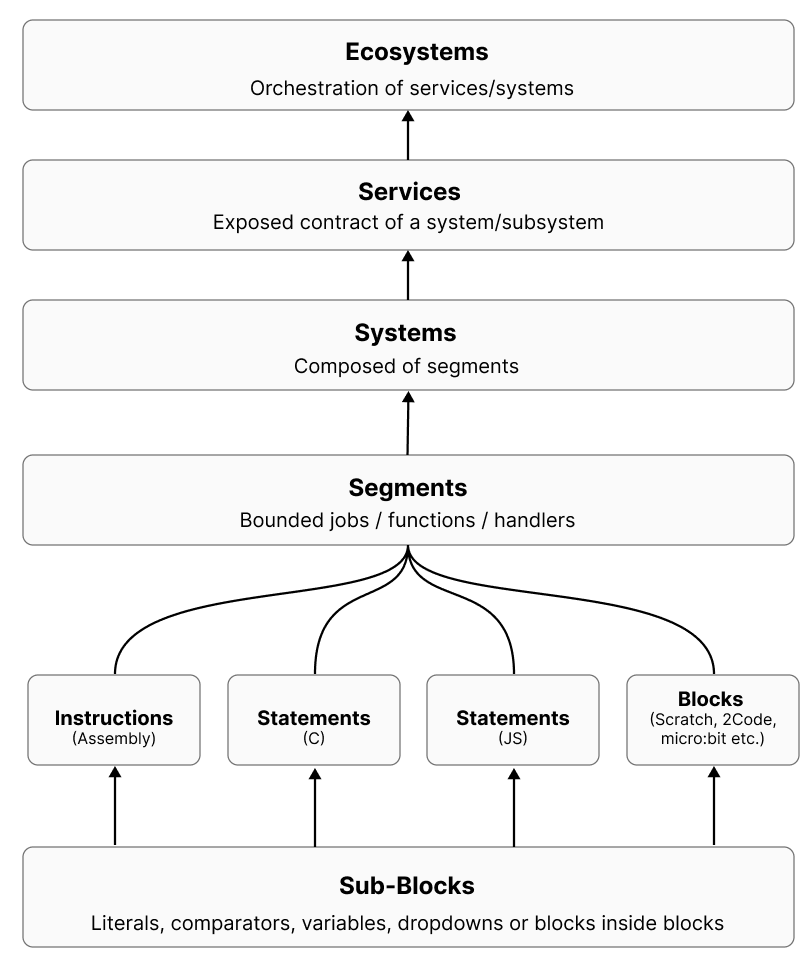}
    \caption{The same element-binding cycle can be applied at each level, but higher layers typically require language- and context-specific overlays (for example, concurrency models, object identity and aliasing, and service-contract concerns).}
    \label{fig:structural-layers}
\end{figure}

Figure~\ref{fig:structural-layers} shows how FLARE~v2's recursive structure can map across different programming paradigms and levels of abstraction. The claim here is not that all paradigms are covered by the present scope, but that the element and binding vocabulary is intended to remain usable as the scale changes, with overlays where required.

\subsection{Binding dimensions: how elements combine}

Section~\ref{sec:diagnosis} identified that FLARE~v1 conflated several distinct mechanisms under a single Relations tier. The design work described in Section~\ref{sec:semiotic} suggested that these mechanisms separate cleanly into two orthogonal dimensions, aligned with Pennington's distinction between control-flow and data-flow abstractions \citep{pennington1987comprehension}.

FLARE~v2 retains Schulte's Block Model insight that comprehension involves structure, flow, and function \citep{schulte2008block}, but refactors these concerns into element properties (Receives, Sends, Effects, Shares) and binding dimensions (Causal-Temporal and Communicative). Two additions distinguish the present framework: recursion, which the Block Model does not formalise; and explicit treatment of Event binding, which accommodates event-driven paradigms where control transfer is governed by hidden runtime mechanisms.

The dimensions are orthogonal in the sense that for any pair of elements, two independent questions can be asked: how does control move between them (Causal-Temporal), and how does information move (Communicative).

\subsubsection{Causal-Temporal binding: how control moves}

The Causal-Temporal dimension describes how execution flows between elements. FLARE~v2 uses three subtypes:

\begin{itemize}
  \item \textbf{Sequential}: No branch instruction is required; the instruction pointer simply increments. One element completes, the next begins. This is the default flow within a segment when no conditional or loop intervenes.

  \item \textbf{Branch}: A jump instruction exists in the user's code - whether for conditionals (\texttt{if}/\texttt{else}), loops (\texttt{while}, \texttt{repeat}), or function calls. The programmer wrote the control structure, and it is visible in their code.

  \item \textbf{Event}: The control transfer exists, but it is governed by a runtime dispatcher (for example an event loop) that is typically hidden from the learner. The learner writes the handler and registration, but not the dispatch mechanism that determines \emph{when} invocation occurs. In common event-loop runtimes, this is often implementable as Branch inside a loop, but from the learner's perspective the dispatch mechanism is not part of the written program and so warrants separate descriptive treatment.
\end{itemize}

\subsubsection{Communicative binding: how data moves}

The Communicative dimension describes how information flows between elements. It operates through three of the four element properties: Sends, Receives, and Shares. (Effects, being treated as private, do not ordinarily participate in inter-element binding.)

\begin{itemize}
  \item \textbf{Sends/Receives}: Transient data flow - return values, parameters, messages passed between elements.
  \item \textbf{Shares}: Persistent state accessible to multiple elements - global variables, hardware state, configuration values, or shared resources.
\end{itemize}

Where the original FLARE~v1 Relations tier bundled "what triggers this" with "what does this share", the two-dimension scheme separates them. Causal-Temporal binding answers \emph{when} and \emph{why} control moves; Communicative binding answers \emph{what} information moves and \emph{where} it persists.

\subsubsection{Why two dimensions, not three?}

An earlier formulation of FLARE~v2 proposed three binding dimensions: Causal-Temporal, Logical-Conditional, and Communicative-Referential. The Logical-Conditional dimension was intended to capture predicates, guards, and Boolean conditions. However, further analysis suggested that conditionals are not a separate \emph{kind} of binding but a mechanism that \emph{governs} branching control flow. The \texttt{if} block, the loop guard, and the Boolean expression are constructs that govern Branch-subtype Causal-Temporal bindings.

Treating conditionals as a separate dimension conflates the mechanism (Boolean evaluation) with the effect (control transfer). The revised scheme places conditionals where they belong: as the means by which Branch bindings are governed. Questions about conditions remain important but are understood as questions about how a Branch binding is controlled, not as a separate binding dimension.

Table~\ref{tab:binding-dimensions} summarises the two dimensions, their subtypes, and representative question stems.


\begin{table}[htbp]
\centering
\caption{Binding dimensions: subtypes, mechanisms, and example question stems}
\label{tab:binding-dimensions}
\small
\begin{tabular}{
P{0.18\linewidth}
P{0.13\linewidth}
P{0.27\linewidth}
P{0.30\linewidth}
}
\toprule
\textbf{Dimension} & \textbf{Subtype} & \textbf{Mechanism} & \textbf{Example question stems} \\
\midrule
\multirow{3}{*}{Causal-Temporal}
  & Sequential & No branch; instruction pointer simply increments. & What happens next? \newline What runs after this? \\
\cmidrule(l){2-4}
  & Branch     & Jump in user code (conditionals, loops, calls).    & Under what conditions does this run? \newline How many times? \\
\cmidrule(l){2-4}
  & Event      & Control transfer governed by hidden runtime dispatch. & What triggers this? \newline When does this fire? \\
\midrule
Communicative
  & (none)     & Data exchange via Sends, Receives, Shares.         & What does this receive? \newline What does it send? \newline What do these share? \\
\bottomrule
\end{tabular}
\end{table}

These dimensions are not proposed as an exhaustive taxonomy; other patterns may matter in paradigms outside the scope stated in Section~\ref{sec:scope}. The claim here is limited: within the imperative and event-driven environments typical of primary and lower-secondary classrooms, these two dimensions and three Causal-Temporal subtypes have been sufficient to support analysis and classroom talk in the author's practice.

\subsection{The recursive element-binding-element cycle}

With elements and binding dimensions established, the core of FLARE~v2 reduces to a four-step recursive cycle:

\begin{enumerate}
  \item \textbf{Identify elements at the current scale.} Decide what counts as an element for the current question.
  \item \textbf{Describe each element using the four properties.} Ask what it receives, sends, effects, and shares.
  \item \textbf{Analyse bindings between the elements.} Examine Causal-Temporal connections (Sequential, Branch, or Event) and Communicative connections (what flows between them, what they share).
  \item \textbf{Recognise the new element that emerges.} Treat the bound set as a new element at the next level, with its own properties, capable of further binding.
\end{enumerate}

This cycle can be applied repeatedly. At one level, it explains how blocks bind into a segment; at the next, how segments bind into a system. In classroom practice, the cycle is enacted through question stems rather than formal steps.

\subsection{Relationship to existing program comprehension models}
\label{sec:positioning}

FLARE~v2 can be situated relative to three established traditions in program comprehension research. It aligns with Pennington's program-versus-situation model, Schulte's Block Model, and Sorva's account of notional machines, while extending each to event-driven environments common in school curricula.

\paragraph{Pennington's dual abstraction.}
Pennington showed that programmers construct two complementary mental representations: a program model that foregrounds control flow, and a situation model that foregrounds data flow and goals \citep{pennington1987comprehension}. FLARE~v2 mirrors this duality in its two binding dimensions: Causal-Temporal binding captures how control moves, while Communicative binding captures how information and shared state move between elements. The extension here is to treat event-driven control as a distinct subtype because its dispatch mechanism is typically hidden from the learner.

\paragraph{Schulte's Block Model.}
The Block Model conceptualises program comprehension as reasoning about multiple hierarchical levels (from atoms to macrostructure) along three dimensions: text surface, execution, and function \citep{schulte2008block}. FLARE~v2 preserves the multi-level insight but makes the level-to-level transition explicit via the recursive element-binding-element cycle, rather than treating levels as a fixed descriptive ladder.

\paragraph{Sorva's notional machines.}
Sorva reviews evidence that learners form mental models of program execution whether or not explicit notional machines are taught, and that hidden runtime mechanisms are a major source of misconception \citep{sorva2013notional}. FLARE~v2 responds by separating visible and hidden control mechanisms within Causal-Temporal binding and by providing a descriptive vocabulary for surfacing hidden dispatch or scheduling in event-driven code.

\FloatBarrier
\section{Illustrative Example: A Simple Control System}
\label{sec:example}

This section illustrates the element-binding-element cycle and the four properties using a minimal three-segment control system commonly encountered in introductory physical computing and robotics. The system, which we refer to as \textit{AutoDimmer}, comprises a \textit{ReadSensor} segment that samples the current light level, a \textit{DecideBrightness} segment that determines an appropriate LED brightness based on a threshold, and a \textit{SetLED} segment that applies the chosen brightness to the hardware. The example is intentionally simple and language-agnostic: its purpose is to make the element-binding-element cycle concrete.

\begin{figure}[H]
    \centering
    \includegraphics[width=0.8\linewidth]{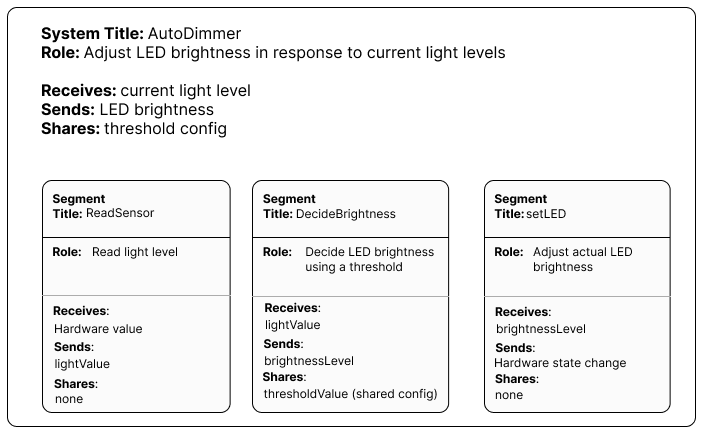}
    \caption{Three segments identified as bounded elements. Their bindings form a simple \newline AutoDimmer control system.}
    \label{fig:segments-system}
\end{figure}

Figure~\ref{fig:segments-system} shows the three segments identified as bounded elements, with their bindings forming the AutoDimmer control system.

\subsection{Step 1: Identify elements at the current scale}

On the first pass through the FLARE cycle, the elements in view are the three segments: \textit{ReadSensor}, \textit{DecideBrightness}, and \textit{SetLED}. In a classroom, a teacher might stabilise these by asking "Which bits of code belong together here?" and then marking or naming each segment as a unit.

\subsection{Step 2: Describe each segment with the four properties}

Each segment can be described using Receives, Sends, Effects, and Shares (Table~\ref{tab:segments-four-properties}). At this stage, questions might include: "What does \textit{ReadSensor} receive and send?", "What does \textit{DecideBrightness} read that is shared?", and "What does \textit{SetLED} write to outside the code?".


\begin{table}[H]
\centering
\small
\caption{Four properties for each segment in the control system}
\label{tab:segments-four-properties}
{%
  \setstretch{1}
  \begin{tabular}{
    P{0.17\linewidth}
    P{0.17\linewidth}
    P{0.17\linewidth}
    P{0.17\linewidth}
    P{0.17\linewidth}
  }
  \toprule
  \textbf{Segment} & \textbf{Receives} & \textbf{Sends} & \textbf{Effects} & \textbf{Shares} \\
  \midrule
  \textit{ReadSensor}        & Trigger from loop or timer & Numeric light level       & None                        & Reads sensor hardware state \\
  \textit{DecideBrightness}  & Light level value          & Chosen brightness value   & Temporary internal variables only & Reads the shared threshold value \\
  \textit{SetLED}            & Brightness value           & Command to LED hardware   & None (no private state)     & Writes to LED device (shared hardware) \\
  \bottomrule
  \end{tabular}%
}
\end{table}

\subsection{Step 3: Analyse bindings between segments}

The next step is to analyse how the segments bind along the two dimensions.

\paragraph{Causal-Temporal binding.}
Sequential binding connects the segments within each cycle: \textit{ReadSensor} runs, then \textit{DecideBrightness}, then \textit{SetLED}. Branch binding operates within \textit{DecideBrightness}, where a conditional compares the sensor reading with the threshold and selects between alternative outputs. The system may also be governed by a timer tick or a loop trigger; depending on environment and representation, that trigger mechanism may be partially hidden from the learner, motivating Event-style questions ("What triggers this whole chain to start each time?").

\paragraph{Communicative binding.}
\textit{ReadSensor} sends the sampled light level and \textit{DecideBrightness} receives it. \textit{DecideBrightness} sends a brightness value and \textit{SetLED} receives it. The threshold value is shared configuration, and the LED is shared hardware state. Any temporary calculations inside \textit{DecideBrightness} are Effects and do not create inter-element bindings, consistent with the definition in Table~\ref{tab:effects-shares}.

\subsection{Step 4: Recognise the system as a new element}

Once the bindings are understood, the three segments can be treated as a single element at the system layer: the \textit{AutoDimmer} system. From this point on, the system can be described using the same four properties (Receives, Sends, Effects, Shares), enabling the cycle to repeat at higher scale.

\section{Discussion: Pedagogical and Curricular Implications}

The Compositional Ladder suggests a progression in which learners first gain fluency with smaller elements and simple bindings, then encounter larger elements whose internal structure can be analysed using the same questions. This is compatible with spiral curriculum principles: concepts are revisited with increasing sophistication, but without introducing an entirely new analytic vocabulary at each stage.

The separation of Causal-Temporal subtypes may be particularly valuable in event-driven environments, where the learner does not see the dispatch mechanism that governs handler invocation. Treating Event as a distinct descriptive subtype enables classroom questions to explicitly surface hidden control, rather than treating all control as if it were user-authored Branch flow. The exploratory Year~5 probe in Section~\ref{sec:semiotic} suggests that early event-handling metaphors may shape pupils' expectations about simultaneous handler activity, reinforcing the value of making hidden dispatch mechanisms explicit in classroom talk.

From a pedagogical perspective, PRIMM \citep{sentance2019primm} provides a temporal sequence for classroom activity (\emph{Predict}, \emph{Run}, \emph{Investigate}, \emph{Modify}, \emph{Make}). FLARE can complement this by providing a consistent descriptive vocabulary for the discussion that occurs within each phase. For example, during \emph{Predict} and \emph{Run}, teachers can stabilise elements and triggers; during \emph{Investigate}, questions can focus on control movement (Causal-Temporal) and information movement (Communicative); and during \emph{Modify} and \emph{Make}, learners can use the same vocabulary to plan what a new element should receive, send, and share. This relationship is a hypothesis for future work rather than an established instructional effect.

In curriculum design, the recursive model is compatible with a staged progression in which new abstraction scales appear once earlier bindings are secure. Early years may emphasise Sequential flow within small elements; later years can introduce Branch bindings and associated conditional reasoning, and then Event bindings and their hidden control mechanisms. Purpose can be introduced at any rung as an interpretive stance ("What is this for? Who relies on it?") without adding another structural tier.

\subsection*{Beyond the system boundary: maintaining continuity with overlays}

A recurring challenge in computing education is the transition from closed programs to open systems, where behaviour depends on environmental factors such as runtime scheduling, service contracts, network latency, permissions, partial failure, and platform policy. In these contexts, a purely code-centred structural description is often insufficient: not because the code loses structure, but because behaviour is increasingly co-produced by the interaction between code and its operating context.

FLARE~v2 accommodates this transition by treating the element-and-binding model as a \emph{structural baseline} and introducing \emph{overlays} as continuity mechanisms above the system layer. Just as Event binding bridges the gap between authored control structures and hidden dispatch, overlays bridge the gap between a system’s internal composition and the external constraints that govern its execution. The learner does not discard structural analysis when moving to system-plus contexts; instead, they extend it.

Practically, this means the element-binding-element cycle remains the first pass for describing the code-side composition, while additional overlays make explicit the environmental forces acting upon that structure (for example, concurrency and scheduling assumptions, service-contract obligations, authority and trust boundaries, or failure semantics). This supports a spiral curriculum: earlier structural reasoning remains valid, but becomes insufficient on its own, and is extended rather than replaced.

\section{Limitations and Future Work}

The Year~5 probe reported in Section~\ref{sec:semiotic} is exploratory and requires systematic replication with controlled conditions. The analysis throughout is practitioner-led and interpretive, drawing on reflective re-reading of earlier classroom work rather than on a newly collected or systematically coded dataset.

Several specific claims require empirical investigation. The assertion that Event binding can often be approximated mechanically as Branch inside a loop but remains descriptively useful for classroom talk raises a testable question: do learners and teachers benefit from this distinction, or does it introduce unnecessary complexity? Similarly, the parallel between literacy progressions and the Compositional Ladder is offered as a potentially useful metaphor for teachers, but whether it reduces cognitive load in practice remains untested.

A further limitation is boundary sensitivity at system-plus. At services and ecosystems, the framework provides a structural baseline for code-side composition and typically needs complementary overlays for operational and socio-technical constraints. Future work should specify, test, and refine a small number of overlays that make these dependencies discussable in classrooms and analysable in research (for example, identity and aliasing for object-based reasoning; explicit concurrency and scheduling models; and service-contract and failure semantics for distributed interaction).

Future research should examine how the recursive binding cycle manifests in teacher and learner discourse. Studies might analyse classroom dialogue using the element-binding-element cycle, investigate how teachers use the Compositional Ladder to structure tasks and questions, or explore whether making hidden runtime control explicit supports more accurate notional machines for event-driven code. Implementation studies and evaluations of classroom uptake will be essential for assessing the practical value of FLARE~v2.

\section{Conclusion}

This paper reframes FLARE from a four-tier descriptive hierarchy (Blocks, Segments, Relations, Macro) to a recursive element-binding-element cycle. The contribution is a compact descriptive scheme for discussing program meaning at increasing scale: bounded elements are described by what they \emph{Receive}, \emph{Send}, \emph{Effect}, and \emph{Share}; and elements bind through two dimensions, Causal-Temporal (Sequential, Branch, Event) and Communicative (Sends/Receives and Shares). The recursive cycle makes explicit how meaning can be composed across levels, rather than treating levels as a fixed classification.

The two-dimension scheme aligns with Pennington's control-flow and data-flow distinction, while accommodating event-driven environments by treating hidden dispatch as a distinct descriptive subtype within Causal-Temporal binding. The scope of the claim remains as stated in Section~\ref{sec:scope}: FLARE~v2 is developed for imperative and event-driven environments typical of primary and lower-secondary curricula. At services and ecosystems, the framework provides a structural baseline for code-side composition and typically needs complementary overlays for operational and socio-technical constraints. This paper does not claim demonstrated coverage of object-oriented design reasoning, explicit concurrency models, distributed systems, or functional paradigms.

FLARE~v2 is presented as a theoretical proposal grounded in classroom need. Its value should be judged by whether it proves generative for analysis, teacher questioning, and curriculum design, and by whether it supports targeted empirical investigation and refinement.

Across these boundaries, the intent is continuity: Event binding preserves structural reasoning where control transfer is mediated by hidden runtime dispatch, and overlays preserve that same reasoning where behaviour is further shaped by environmental constraints beyond the system boundary.

\section{Acknowledgements}

With thanks to Robert Whyte and Sue Sentance, whose guidance and collaboration during the TICE2 project and CEP~'25 study laid the foundations for the ongoing development of FLARE.

\section{Funding}

The author participates in the Microsoft for Startups Founders Hub programme, which provided Azure credits used for technical prototyping of a tool that applies the FLARE framework. No external funding supported the research reported in this paper.

\section{Conflict of Interest}

The author is developing a tool that applies the FLARE framework, which may be commercialised in the future. The work reported in this paper was completed independently of the tool's development. No other conflicts of interest are declared.

\newpage

\bibliographystyle{ACM-Reference-Format}
\bibliography{paper_bib}

@inproceedings{heath2025flare,
  author    = {Heath, Justin and Whyte, Robert and Sentance, Sue},
  title     = {{FLARE}: A Framework Supporting Code Comprehension and Formative Assessment in Block-Based Programming Education},
  booktitle = {Proceedings of the 9th Conference on Computing Education Practice},
  series    = {CEP '25},
  year      = {2025},
  pages     = {25--28},
  publisher = {Association for Computing Machinery},
  address   = {New York, NY, USA},
  isbn      = {9798400711725},
  url       = {https://doi.org/10.1145/3702212.3702219},
  doi       = {10.1145/3702212.3702219}
}

@article{black1998assessment,
  author  = {Black, Paul and Wiliam, Dylan},
  title   = {Inside the Black Box: Raising Standards through Classroom Assessment},
  journal = {Phi Delta Kappan},
  year    = {1998},
  volume  = {80},
  number  = {2},
  pages   = {139--148},
  url     = {https://www.jstor.org/stable/20439383}
}

@incollection{sentance2023programming,
  author    = {Sentance, Sue and Waite, Jane},
  title     = {Programming in the classroom},
  booktitle = {Computer Science Education: Perspectives on Teaching and Learning in School},
  editor    = {Sentance, Sue and Barendsen, Erik and Howard, Nicol R. and Schulte, Carsten},
  publisher = {Bloomsbury Academic},
  address   = {London},
  year      = {2023},
  pages     = {275--290}
}

@book{hjelmslev1961prolegomena,
  author    = {Hjelmslev, Louis},
  title     = {Prolegomena to a Theory of Language},
  year      = {1961},
  publisher = {University of Wisconsin Press},
  address   = {Madison, WI},
  note      = {Original work published 1943, translated by F. Whitfield}
}

@incollection{pennington1987comprehension,
  author    = {Pennington, Nancy},
  title     = {Comprehension strategies in programming},
  booktitle = {Empirical Studies of Programmers: Second Workshop},
  editor    = {Olson, Gary M. and Sheppard, Sylvia and Soloway, Elliot},
  pages     = {100--113},
  year      = {1987},
  publisher = {Ablex},
  address   = {Norwood, NJ}
}

@techreport {royalsociety2017,
  author = {{The Royal Society}},
  title  = {After the Reboot: Computing Education in UK Schools},
  year   = {2017},
  institution  = {The Royal Society},
  address      = {London},
  type         = {Report},
  url    = {https://royalsociety.org/~/media/policy/projects/computing-education/computing-education-report.pdf}
}

@techreport{royalsociety2025,
  title        = {System upgrade required: Creating opportunities in computing education},
  author       = {{The Royal Society}},
  year         = {2025},
  month        = {October},
  institution  = {The Royal Society},
  address      = {London},
  type         = {Report},
  url          = {https://royalsociety.org/-/media/education/computing-in-schools/system-upgrade-required-report.pdf}
}

@book{saussure1916course,
  author    = {de Saussure, Ferdinand},
  title     = {Course in General Linguistics},
  year      = {2011},
  publisher = {Columbia University Press},
  address   = {New York, NY},
  note      = {Edited by P. Meisel and H. Saussy, translated by W. Baskin. Original work published 1916}
}

@inproceedings{schulte2010block,
  author    = {Schulte, Carsten and Clear, Terry and Taherkhani, Alireza and Busjahn, Timo and Paterson, John H.},
  title     = {An Introduction to Program Comprehension for Computer Science Educators},
  booktitle = {Proceedings of the 2010 ITiCSE Working Group Reports (ITiCSE-WGR '10)},
  pages     = {65--86},
  year      = {2010},
  publisher = {ACM},
  address   = {New York, NY, USA},
  doi       = {10.1145/1971681.1971687}
}

@article{sentance2019primm,
  author = {Sentance, Sue and Waite, Jane and Kallia, Maria},
  title = {Teaching Computer Programming with {PRIMM}: A Sociocultural Perspective},
  journal = {Computer Science Education},
  volume = {29},
  number = {2--3},
  pages = {136--176},
  year = {2019},
  doi = {10.1080/08993408.2019.1608781}
}

@article{sorva2013notional,
  author  = {Sorva, Juha},
  title   = {Notional Machines and Introductory Programming Education},
  journal = {ACM Transactions on Computing Education},
  year    = {2013},
  volume  = {13},
  number  = {2},
  pages   = {8:1--8:31},
  doi     = {10.1145/2483710.2483713}
}

@inproceedings{schulte2008block,
  author = {Schulte, Carsten},
  title = {Block Model: An Educational Model of Program Comprehension as a Tool for a Scholarly Approach to Teaching},
  booktitle = {Proceedings of the Fourth International Workshop on Computing Education Research (ICER '08)},
  year = {2008},
  pages = {149--160},
  publisher = {ACM},
  address = {New York, NY}
}

@misc{heath2025concurrency,
  author = {Heath, Justin},
  title = {When Year 5 Children Reason Better About Concurrency than Our Theories Predict},
  howpublished = {Substack},
  year = {2025},
  month = dec,
  note = {\url{https://open.substack.com/pub/justinquisitive/p/when-year-5-children-reason-better}},
}

@article{soloway1984empirical,
  author = {Soloway, Elliot and Ehrlich, Kate},
  title = {Empirical Studies of Programming Knowledge},
  journal = {IEEE Transactions on Software Engineering},
  volume = {SE-10},
  number = {5},
  pages = {595--609},
  year = {1984}
}

\bigskip

\begin{singlespace}

\section*{Licence}

This preprint is made available under the Creative Commons Attribution-NonCommercial-ShareAlike 4.0 International Licence (CC~BY-NC-SA~4.0). This permits copying, distribution, and adaptation for non-commercial purposes, provided that appropriate credit is given and that derivative works are released under the same licence.

This licence applies only to the text, figures, and explanations contained in this paper. Practical implementations of the framework - including training materials, classroom scaffolds, exemplar resources, and software tools - are licensed separately and are not covered by this CC BY-NC-SA 4.0 licence.

\paragraph{Author's intention.}
FLARE is released openly to support teachers' professional judgement rather than prescribe scripted schemes of work. The intention is to strengthen educator autonomy, not to constrain it.

\end{singlespace}

\end{document}